# Titanium dioxide synthesized using titanium chloride: Size effect study using Raman and Photoluminescence

S. K. Gupta, Rucha Desai, P. K. Jha[$], S. P. Sahoo[*] and D. Kirin[#]

Department of Physics, Bhavnagar University, Bhavnagar, 364 022, Gujarat, India
[*]Material Science Division, Indira Gandhi Centre for Atomic Research, Kalpakkam, 603102, India
[#]Ruđer Bošković Institute Bijenička Cesta 54, P.O.B. 180, HR-10002 Zagreb, Croatia

*Abstract*

Titanium dioxide ($TiO_2$) nanocrystals were prepared by wet chemical method and characterized by x-ray diffraction (XRD), transmission electron microscopy (TEM), Raman scattering (RS) and photoluminescence (PL). The X-ray diffraction shows the formation of nanocrytalline $TiO_2$ of average sizes ~7 nm and ~15 nm for two samples. The x-ray diffraction, transmission electron microscopy (TEM) and Raman scattering shows that the $TiO_2$ nanocrystals has anatase crystal structure for both samples. The PL intensity of the smaller particle is more, which has been attributed to defects and particle size variation. A modified phonon confinement model with the inclusion of size distribution, a new confinement function for $TiO_2$ nanocrystals and averaged dispersion curves for most dispersion phonon branch ($\Gamma$-$X$ direction) has been used to interpret the size variation of Raman spectra. The obtained Raman peak shift and FWHM agree will the experimental data. Our observations suggest that phonon confinement effects are responsible for a significant shift and broadening for the Raman peaks.

# INTRODUCTION

Titanium dioxide ($TiO_2$) is a wide band gap semiconductor, crystalline as anatase, rutile or brookite and is highly attractive material for variety of industrial applications such as solar cells, photocatalysis, charge spreading devices, chemical sensors, microelectronics, electrochemistry, and dye sensitized solar cell etc [1]. Another imperative feature in the development of a way to use sunlight, which represents a valuable source of renewable energy, since only small UV fraction of solar light can be utilized due large bandgap of 3.2 eV in the anatase crystalline phase [2, 3]. However, the challenge for so called nanotechnologies is to achieve perfect control of nanoscale-related properties. This requires correlating the parameters of the synthesis process with the resulting nanostructure. It is important to develop synthesis methods in which the morphology and the structure of nano titania can be controlled or the fine structure can be mimicked at the nanoscale, owing to its technological importance, which is due to its strong oxidizing power, chemical inertness, and non-toxicity. The $TiO_2$ nanocrystals for which the anatase phase is most common have recently received attention due to its very interesting properties, different from the bulk and hence they are not only important from the application but also from the fundamental of view. Further, the $TiO_2$ nanocrystal in its anatase phase has higher photocatalytic activity than other phases and its bulk counterpart [4]. The anatase nanocrystals is easily prepared in a range of sizes, shapes, and forms (as nanocrystals, rods, wires, etc.) by using a variety of physical and chemical methods. Synthesis route of $TiO_2$ production usually result in amorphous solid $TiO_2$ or anatase or other phase depending on the preparation routes and the experimental conditions. However, a method which provides the control size distribution, crystalline phase and stoichiometry is essential for the application of $TiO_2$ nanoparticles.

Synthesis and characterization of $TiO_2$ nanoparticle is state of art, the properties depend on how we synthesize and characterize the materials. The shape and crystal structure of the titanium dioxide product significantly depends on the particle size of the titanium used as a raw material. The anatase phase is thermodynamically more stable than rutile phase at sizes < 14 nm, but marginally more stable than brookite at sizes <~11 nm [5, 6]. Therefore, the synthesis of nanotitania in single phase anatase is a challenge as the brookite is likely to be present with anatase. Pottier *et al* [7] synthesized nanocrytalline $TiO_2$ by precipitation routes, Kim *et al* [8] by microemlusion techniques, Music *et al* [9] and Bersani *et al* [10] by sol-gel methods, Melendres *et al* [11] used physical or chemical vapor deposition technique and Tang *et al* [12] synthesize by organometallic routes. Frequently hydrolysis of alkoxides is used to produce $TiO_2$ nanopowders. Their synthesis was performed in a sealed glass ampoule at various temperatures, the crystallite size was controlled by the reaction temperature but needs a post thermal annealing treatment to obtain the desired crystalline phase. Further the size distribution was not narrow and difficult to produce in bulk.

Wet chemical route [13] is a suitable method to control the particle size, shape, size distribution and crystalline phase. The chemical route for the production of nanocrystals allows excellent control over the mean particle size, particle size distribution, transition temperature between the different phases, physical and mechanical properties.

The Raman spectroscopy and Photoluminescence is a non destructive technique and has been widely used to characterize the $TiO_2$ nanocrystals and understand its size dependent phonon properties in anatase nanocrystals $TiO_2$. The Raman spectroscopy also allows obtaining the role of strain energy, quantization effect and formation of stable phase of $TiO_2$ nanoparticle. In the study of the nanocrystalline anatase $TiO_2$ by Raman spectroscopy, a shift in the peak position to a higher energy side and a broadening in full width at half maximum (FWHM) of the main anatase mode (144 $E_g$) have been reported [14-17]. These observations have been attributed to either phonon confinement (PC) [18-20] or nonstoichiometry or internal stress in nanocrystals. Size-dependent shift and linewidth in the Raman spectra of the anatase nanocrystals have been explained by phenomenological PC models [18-21] by various researchers. However, none of these phonon confinement models [18-21] gives a satisfactory description of the phonon within the dot, capable of determining simultaneously the frequency shift and the linewidth. Most of the models based on phenomenologically theory give the qualitative description of the size dependent Raman shift, use arbitrary confinement function and dispersion curves. Furthermore, poor agreement between experimental data and theoretical calculation may lead to diverge the subject. However, the nonsystematic Raman spectral modifications that are inconsistent with phonon confinement [21-23] documented for the nanoscale rutile within thin films suggest that the intrinsic crystallographic or extrinsic matrix influences can interfere and overprint the confinement effects. To interpret the size variation of Raman spectra, spatial correlation models so far use, the randomly selected Gaussian, exponential and sine functions so far have been used as confinement function without any proper justification. The dispersion curves used in the model are considered from its bulk dispersion curves are also arbitrarily selected. Therefore a phonon confinement model with improved expression, which includes the particle size distribution, proper confinement function and dispersion curves, allowing the determination of the frequency shift and of the line broadening in Raman spectra , as a function of the nanoparticles size is the need of the time.

Photoluminescence (PL) spectroscopy is another powerful tool to obtain information about the electro-optic and photoelectric properties of nanomaterials as it depend on electronic excitations. PL gives the possibilities to study the key question in quantum dots (QDs) spectroscopy deals with the determination of the discrete energy spectrum. The PL study of the $TiO_2$ nanocrystals attributes different mechanism (self-trapped excitons, oxygen vacancies and surface state etc.) to the size dependent PL spectra of anatase $TiO_2$ nanocrystal [24, 25].

Despite the fact that the tetragonal form of $TiO_2$ has dielectric, optical, and elastic properties of considerable interest no serious attention has been paid for the systematic characterizations of the anatase $TiO_2$ nanocrystals particularly by using the proper model to interpret the Raman spectra and PL spectra. This may be due to the difficulty of producing anatase nanoparticles with a controlled stoichiometry, mass distribution and without contaminations [26- 28]. The present study is aimed to synthesize a single phase anatase $TiO_2$ nanocrystal by the wet chemical route and study the size dependent shift in the peak and linewidth in its Raman and PL spectra. An improved expression for phonon confinement which includes the new confinement function, dispersion curves and log-normal size distribution obtained from the TEM measurement is used to analyze the Raman spectra. Size of the $TiO_2$ nanoparticle is obtained from XRD and TEM images and compared with the size obtained from Raman and PL spectra.

**EXPERIMENTAL DETAILS**

The titanium dioxide nanoparticles were synthesized using wet chemical technique. The analytical grade titanium tetrachloride ($TiCl_4$) was used. Other chemicals used were analytical grade ammonium hydroxide solution (35%), ethylene glycol, polyethylene glycol and sodium acetate. With vigorous stirring titanium chloride (2 ml) was drop wise mixed in ammonium hydroxide/ethylene glycol and continued stirred for 10 mins. The reaction was exothermic and carried out in 100 ml beaker. In method-I, this mixed precursor was heated to 333K. White particles were washed several times using warm water to remove chlorine impurities. Particles were dried by applying acetone wash. As-prepared particles showed amorphous phase. In order to convert it into crystalline phase, as-prepared particles were heated to 623 K. In method-II, once transparent solution was obtained sodium acetate (3.6 g) and polyethylene glycol (0.9 ml) were added and vigorously stirred for 30 mins. This mixture was heated to 423 K in closed system. Once the condensation starts white color precipitate formed, this process was continued for 4 hours. The white colored particles were collected using centrifugation. The particles were washed using warm distilled water followed by acetone wash. The as-prepared particles were annealed at 573 K for 1 hr. The samples obtained using method I and II are designated here as A and B respectively. These samples were characterized using Powder X-ray diffractometer (Bruker) to understand the structural characteristics. To study the effect of particle size on the Raman spectra and Photoluminescence (PL) spectra experiments were performed using U1000 and Spex-14018 Raman spectrometer with a backscattering geometry at room temperature with 532 nm line of DPSS laser and 457.9 nm line of an argon-ion laser respectively. However, we have also recorded the micro Raman spectra of powder samples on Horiba Jobin Yuon T64000 spectrometer, Argon ion laser COHERENT INNOVA 400 operating at 514.5 nm, with the laser power ranging from 10-100 mW. We did not observe any change in the spectra except a slight change in Intensity. Transmission electron microscopy (TEM) was performed by using JEOL JEM 2100 200 kV microscope, having point resolution is 50 X to 1.5 MX.

**PHONON CONFINEMENT EFFECT**

The role of dimensional confinement in modifying acoustic phonon modes and their interactions with charge carriers plays important role in the physics of nanostructures. As the size of nanocrystal decreases, phonons with a larger wave vector are involved in the electron-phonon interaction and hence the electron-phonon interaction with acoustic phonon modes becomes more pronounced as compared to the scattering of electrons by optical phonons. Furthermore, in nanostructures, dimensional confinement is arising due to the phase space restrictions, which may weaken or forbid for the optical phonon scattering processes normally dominate in bulk structures.

In a perfect crystal the first-order Raman scattering of a photons selects contributions of phonons obeying the equation $q \approx 0$ ($q$ is the phonon wave vector), because of the momentum conservation law. Therefore, this selection rule implies the contribution of phonons at the center of the Brillouin zone. However, when the size of the crystal is quite limited, the $q \approx 0$ is no longer valid and $q \neq 0$, contribution of phonons (away from zone centre) also contribute to the Raman lineshape. Consequently, the Raman spectrum [18-23, 29-30] is calculated by the following integral over the momentum vector **q**,

$$I(\omega) \propto [n(\omega) + 1] \int |C(q)|^2 L(\omega, q) dq \tag{1}$$

Where the integral is extended to the entire first Brillouin zone, $\omega$ is the Raman frequency, $n+1$ is the Bose-Einstein factor, $C(q)$ are Fourier coefficients of the phonon wave function and $L(\omega, q)$ is the Lorentzian function related to the phonon dispersion curve. For low-dimensional nanocrystals, the Raman line shape expressed as [30],

$$I(\omega) \alpha \int \rho(L) dL \int \left( \frac{|c(q)|^2}{(\omega - \omega(q))^2 + (\Gamma_0)^2} \right) d^3q \tag{2}$$

Where $\Gamma_0$ is the FWHM of the Raman line of bulk, $\omega(q)$ is the phonon dispersion curve, $c(q)$ is the Fourier coefficients of the vibrational weighting function expanded in the Fourier integral and $\rho(L)$ is the log-normal distribution. Initial wave vector $k_n$ can contribute to the photon scattering process with an effective wave vector **q** with the spread allowed by the Heisenberg uncertainty principle around the initial value of $k_n$. It can be written as $k_n = n\pi/D$ with D as the size of spherical nanocrystal and log-normal distribution $\rho(L)$ can be defined as,

$$\rho(L) = Exp\left(\frac{-Log^2\left(\frac{D}{D_0}\right)}{2\sigma^2}\right) \quad (3)$$

$D_0$ is the distribution for size of spherical nanocrystals and $\sigma$ is the size distribution. The confinement function $c(q)$ for spherical particles in present case is considered as [18, 23, 31],

$$|c(q)|^2 \approx \left(\frac{3 Sin\left(\frac{qD}{2}\right)}{\left(\pi^3 D^3 q - (k_n^2 - q^2)\right)}\right)^2 \quad (4)$$

In the calculation of Raman intensity, we have used $\Gamma_0 = 7.5$ cm$^{-1}$ and $\omega = 143$ cm$^{-1}$. The size distribution of the particles and dispersion curves, $\omega(q)$ in the calculation of Raman intensity are very important for the description of phonon confinement in nanocrystals [21] which have been ignored in most of the phonon confinement models. Here in the present calculation, we consider both. The size distribution is considered as log-normal distribution presented by the expression (3). Most of the earlier cases the dispersion curves are considered one which is easy to be presented or fit to the smoother curves [29]. In the present calculation we have calculated the Raman intensity for the two most variant dispersion curves of the phonon branch in **Γ-X** direction. However, we present the averaged peak shift and FWHM obtained from two dispersion curves. In the below we discuss the some of the earlier used dispersion curves and present considered dispersion curves for the calculation of Raman Intensity, peak shift and FWHM. Bersani *et al* [15] used the following dispersion relation,

$$\omega(q) = 142.5 + 20(1 - Cos(qa)) \quad (5)$$

Where *a* is 0.3768 nm. Ivanda *et al* [30] used a weighted average of phonon dispersion,

$$\omega(q) = 142.5 + 164 Sin^2(qx) \quad (6)$$

Here $x$= 1.5177. Balaji *et al* [29] have taken the dispersion curves for 143 cm$^{-1}$ branch from the phonon dispersion curves from Mikami's [32] work and accommodated by computing from each of these two modes (**Γ-N**) as averaged from individual phonon modes. However the dispersion curves along **Γ-X** is neglected due to its very high degree of variation. In the present case the two dispersion curves along **Γ-X** have been extracted from Mikami's [32] work and their expressions are presented below,

$$\omega(q) = 142.5 + 107(1 - Cos(3.0\, q^{0.745})) \text{ and} \tag{7}$$

$$\omega(q) = A + B_1(q\pi) + B_2(q\pi)^2 + B_3(q\pi)^3 + B_4(q\pi)^4, \tag{8}$$

with A=141.6 cm$^{-1}$, B$_1$=121.9 cm$^{-1}$, B$_2$=-820.9 cm$^{-1}$, B$_3$=1536.7 cm$^{-1}$, B$_4$=-781.8 cm$^{-1}$.

**RESULTS AND DISCUSSION**

Figure 1 shows the X-ray diffraction pattern of TiO$_2$ nanoparticles samples A and B. It is evident from the figure that both samples possess the single phase anatase structure. No lines correspond to any secondary and/or impurity phase(s) are observed. In figure 1 the diffraction lines are indexed for anatase structure. It is evident from the figure that the intensity of all peaks increases in sample B compared to A. This increase of intensity suggests increase of particle size and/or better crystalline structure. The average particle size calculated using Scherer's formula for the most intense (101) peak for samples A and B are ~7 and ~15 nm respectively. Figure 2 shows TEM image captured by suspending the particle into ethanol for sample A. Selected area diffraction (SAD) pattern is shown in inset of figure 2. The formation of ring pattern with the dotted line indicates the highly crystalline sample. The TEM image reveals that particles are spherical or going to be spherical in shape. Particle size histogram calculated from 2 to 20 particles is shown in inset of figure 2. It is clear that main contribution is from ~8 nm sized particles and is in good comparison with XRD result.

Figure 3 presents the Raman spectra for two different sizes of $TiO_2$ nanoparticles (samples A and B). The spectra are typical of the anatase $TiO_2$ phase confirm the phase obtained from XRD. The first $E_g$ peak at ~144 cm$^{-1}$, a characteristic of anatase $TiO_2$ is slightly broader and shifted with respected to those of a bulk $TiO_2$ crystal. The broadening and blue shift of the Raman peaks observed in $TiO_2$ nanocrystals are attributed to either surface pressure or phonon confinement effect that usually exist in nanometer-sized materials [21-22]. According to group theoretical analysis there are six Raman active modes for tetragonal anatase structure with two formula units per unit cell. The Raman peaks observed at 143, 197, 397, 516 and 637 cm$^{-1}$ in Fig. 3 are assigned as the $E_g$, $E_g$, $B_{1g}$, $A_{1g} + B_{1g}$, and $E_g$ modes of anatase phase, respectively. The occurrence of intense 143 cm$^{-1}$ mode indicates that, ~7 nm $TiO_2$ nanocrystals possess a certain degree of long-range order, whereas the weak broader peak in the high-frequency region indicates that there is lack of short-range order in the anatase phase [15, 24]. The second $E_g$ peak becomes more prominent in the case of 7 nm $TiO_2$ nanocrystals. The shift in the Raman peak which has been attributed earlier to the surface pressure [21-22] is not due to surface pressure in the present $TiO_2$ nanocrystals as they are free and not embedded in any medium but the phonon confinement is the main cause as will be seen in what follows. The first $E_g$ mode is influenced by the grain size of the $TiO_2$ nanocrystals, which shows blue shifts in frequency and increase in linewidth with a decrease of grain size. Figure 4 presents the size variation of Raman Intensity calculated by the Eqn. (2) for the first $E_g$ mode and reveals the blue shift of the peak position with decreasing size of the $TiO_2$ nanoparticles. Figure 5 shows the peak position and Full width and half maximum (FWHM) of the main anatase $E_g$ Raman mode as a function of the nanoparticle size calculated using the dispersion relation extracted from the density functional perturbation theory calculations of Mikami *et al* [32], new confinement function and particle size distribution, which was ignore in Bersani [15], Ivanda [30], and Balaji *et al* [29]. The two dispersion curves for the **Γ–X** direction which have been considered in the present calculations are presented in the inset of figure 5 (b). It is seen from fig. 5 that the FWHM and peak position obtained from weighted average intensity fit the crystallite size obtained from experimental data (which is shown in fig. 5), within the experimental error limits and provide a good comparison form smaller particles. The figure 5 also includes the earlier calculated data. It can also seen from fig. 5 that FWHM well matched with the experimental data, which may be due to the particle size distribution included in the phonon confinement function. The peak position is well matched for smaller size, due to high dispersion curve and particle size distribution and diverse for bigger particles.

Figure 6 depicts the room temperature (PL) photoluminescence emission spectra for the $TiO_2$ nanocrystals of samples A (~7 nm) and B (~15 nm) recorded at 457.9 nm wavelength range. These samples gave the similar emission spectra with broadened peaks at 2.49 eV. However, there was a small blue shift in the case of smaller particle. It is obvious that decrease in particle size increase the surface energy, which can be attributed to the quantum confinement, consequently blue shift (≈0.053 eV) in emission peak is observed. The increase in PL intensity for ~7 nm sample is due to the self trapped excitons recombination which is a combined effect of defect centers generated from oxygen vacancies and particle size. This is in contrast to the recent study on the PL spectra of $TiO_2$. A recent report on the PL spectra analysis of anatase $TiO_2$ nanoparticle [24] also exhibited the UV luminescence centered at 2.24 eV and 2.16 eV are in agreement with the present report. The present result on the PL spectra is similar to the CdS and CuCl nanocrystals, when the spectra intensity was enhanced and the maximum of the emission band was blueshifted [33, 34]. Hence in the present case the PL is attributed to both to confinement and defect PL.

**CONCLUSION**

In summary, the TiO$_2$ nanocrystals have been synthesized through a cost effective chemical route method. XRD, TEM and Raman results confirm a single phase anatase TiO$_2$ nanocrystal. The phonon confinement was found the main cause of the broadening and blue shift of the main Raman peak in anatase nanocrystals. The modified phonon confinement model considered in the present approach is able to produce the size variation of Raman peak shift and FWHM better than the previous calculations and hence confirms the importance of proper confinement function and dispersion curves along with the particle size distribution in the presentation of Raman Intensity by using the Phonon confinement models. Sizes of the particles obtained from TEM, XRD, Raman and PL are consistent. The increase in PL intensity for smaller nanoparticle (~7 nm) is due to the self trapped excitons recombination which is a combined effect of defect centers generated from oxygen vacancies and particle size. There is a slight blue shift in the photoluminescence peak with decrease in the size arising due to the increase in surface energy.

**ACKNOWLEDGEMENTS**

It is a pleasure to acknowledge Dr. A K Arora for discussion and Dr. K. Furić to help in recording the micro-Raman spectra at IRB, Zagreb. The financial assistance from DAE-BRNS and DST are highly appreciated.

**REFERENCES**

1. M. Gratzel, *Nature* (London) **414**, 338 (2001).

2. X. Chen and S. S. Mao, *Chem. Rev.* (Washington, D. C.) **107**, 2891 (2007).

3. Q. Xu and M. A. Anderson, *J. Am. Ceram. Soc.* **77**, 1939 (1994).

4. L. Gao and Q. Zhang, Scr. Mater. **44**, 1195 (2001).

5. H. Zhang and J. F. Banfield, J. Phys. Chem. B **104**, 3481 (2000).

6. H. Zhang and J. F. Banfield, J. Mater. Chem. **8**, 2073 (1998).

7. A. Pottier, S. Cassaignon, C. Chanéac, F. Villain, E. Tronc and J. Jolivet, *J. Mater. Chem* **13**, 877 (2003).

8. E. J. Kim and S. H. Hahn, *Mater. Lett.* **49**, 244 (2001).

9. S. Musić, M. Gotić, M. Ivanda, S. Popović, A. Turković, R. Trojko, A. Sekulić and K. Furić, *Mater. Sc. Eng.* B **47**, 33 (1997).

10. D. Bersani, G. Antonioli, P. P. Lottici and T. Lopez, *J. Non-Cryst. Sol.* **232**, 175 (1998).

11. A. Melendres, A. Narayanasamy, V. Maroni and R. Siegel, *J. Mater. Res*. **4**, 1246 (1989).

12. J. Tang, F. Redl, Y. Zhu, T. Siegrist, L. E. Brus and M. L. Steigerwald, *Nano Lett*. **5**, 543 (2005).

13. Y. Chen, A. Lin and F. Gan, *Powder Techn.* **167**, 109 (2006).

14. G. J. Exarhous, N. J. Hess, *Thin Solid Films* **220**, 254 (1992).

15. D. Bersani, P. P. Lottici and X. Z. Ding, *Appl. Phys. Lett*. **72**, 73 (1997).

16. V. Swamy, A. Kuznetsov, L. S. Durbrovinsky, R. A. Caruso, D. G. Shchukin, B. C. Muddle, *Phys. Rev*. B **71**, 184302 (2005).

17. K.-R, Zhu, M.-S. Zhang, Q. Chen, Z. Yin, *Phys. Lett* A **340**, 220 (2005).

18. H. Richter, Z. P. wang, L. Ley, *Solid State Commun*. **39**, 625 (1981).

19. P. M. Fauchet, I. H. Campbell, *Crit. Rev. Solid State Mater. Sci.* **14**, S79 (1988).

20. I. H. Campbell, P. M. Fauchet, *Solid State Commun*. **58**, 739 (1986).

21. A. K. Arora, M. Rajalakshmi, T. R. Ravindran, and V. Sivasubramanian, *J. Raman Spectrosc*. **38** 604 (2007).

22. A. K. Arora, M. Rajalakshmi, and T. R. Ravindran in *Encyclopedia of Nanoscience and Nanotechnology* Vol. 8, Ed. H.S. Nalwa (American Scientific Publishers, Los Angeles, 2004) p.499.

23. P. K. Jha and M. Talati, *Phonons in nanocrystal quantum dots*, Focus on Condensed Matter Physics, Chap. 2, pp. 83-103, Ed. John V. Chang, Nova Science Publishers, NY (2006).

24. W. F. Zhang, M. S. Zhang, Z. Yin, Q. Chen, *Appl. Phys*. B **70**, 261 (2005).

25. M. Rajalakshmi, A. K. Arora, B. S. Bendre and S. Mahamuni, J. Appl. Phys. **87,** 2445 (2000); M. Rajlakshmi and A. K. Arora, Nanostruct. Materials, **11**, 399 (1999).

26. T. Mazza, E. Barborini, P. Piseri and P. Milani, *Phys. Rev.* B **75**, 045416 (2007).

27. V. Swamy, *Phys. Rev*. B **77**, 195414 (2008).

28. V. Swamy, A. Kuznetsov, L. S. Dubrovinsky, P. F. McMillan, V. B. Prakapenka, G. Shen and B. C. Muddle, *Phys. Rev. Lett*. **96**, 135702 (2006).

29. S. Balaji, Y. Djaoued and J. Robichaud, *J. Raman Spectosc.* **37**, 1416 (2006).

30. M. Ivanda, S. Music, M. Gotic, A. Turkovic, A. M. Tonejc, O. Gamulin, *J. Mol. Struct*. **480**, 641 (1999).

31. V. Paillard, P. Puech, M. A. Laguna, R. Carles, B. Kohn and F. Huisken, *J. Appl. Phys.* **86**, 1921 (1999).

32. M. Mikami, S. Nakamura, O. Kitao, H. Arakawa, *Phys. Rev*. B **66**, 155213 (2002).

33. R. Rossetti, J. L. Ellison, J. M. Gibson, L. E. Brus, J. Chem. Phys. **80**, 4464 (1984).

34. T. Itoh, M. Furumiya, T. Ikehara, C. Gourdon, Solid State Commun. **73**, 271 (1990).

**Figure Caption**

**Figure 1:** XRD spectra of sample A and B.

**Figure 2:** Transmission electron microscopy (TEM) image of sample A, inset with particle size distribution and selected area diffraction (SAD) pattern of same sample.

**Figure 3:** Raman spectra of $TiO_2$ powders with different nanocrystals size. The mode symmetries of the anatase are indicated (Inset figure shows first $E_g$ mode for both sample A and B).

**Figure 4:** Calculated Raman intensity of the first $E_g$ mode with size of the $TiO_2$ nanoparticles.

**Figure 5:** Calculated FWHM (a) and peak shift (b) of Raman line shape of anatase TiO2 as a function of particle size. Various symbols indicate the experimentally observed and dotted curve, dashed curves shows the FWHM and peak position *vs* crystallite size performed by using different dispersion relations. Inset figure shows dispersion curves.

**Figure 6:** PL spectra of $TiO_2$ powders with different nanocrystals size.

**Figure 1**

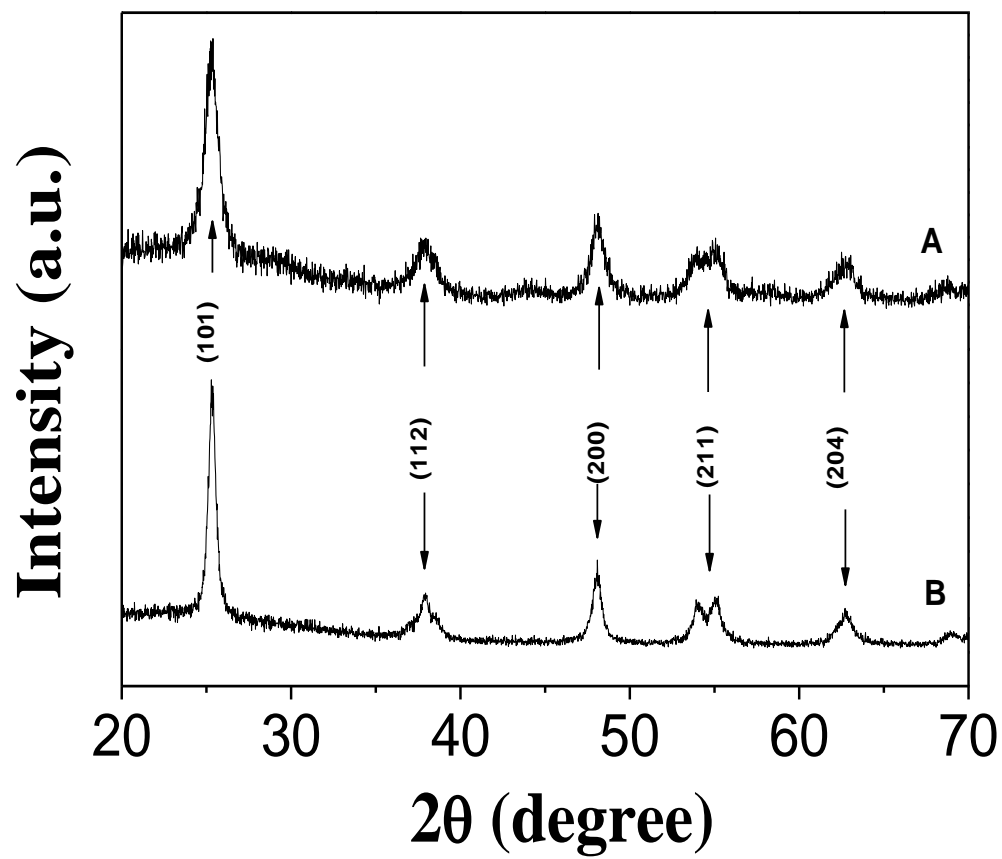

**Figure 2**

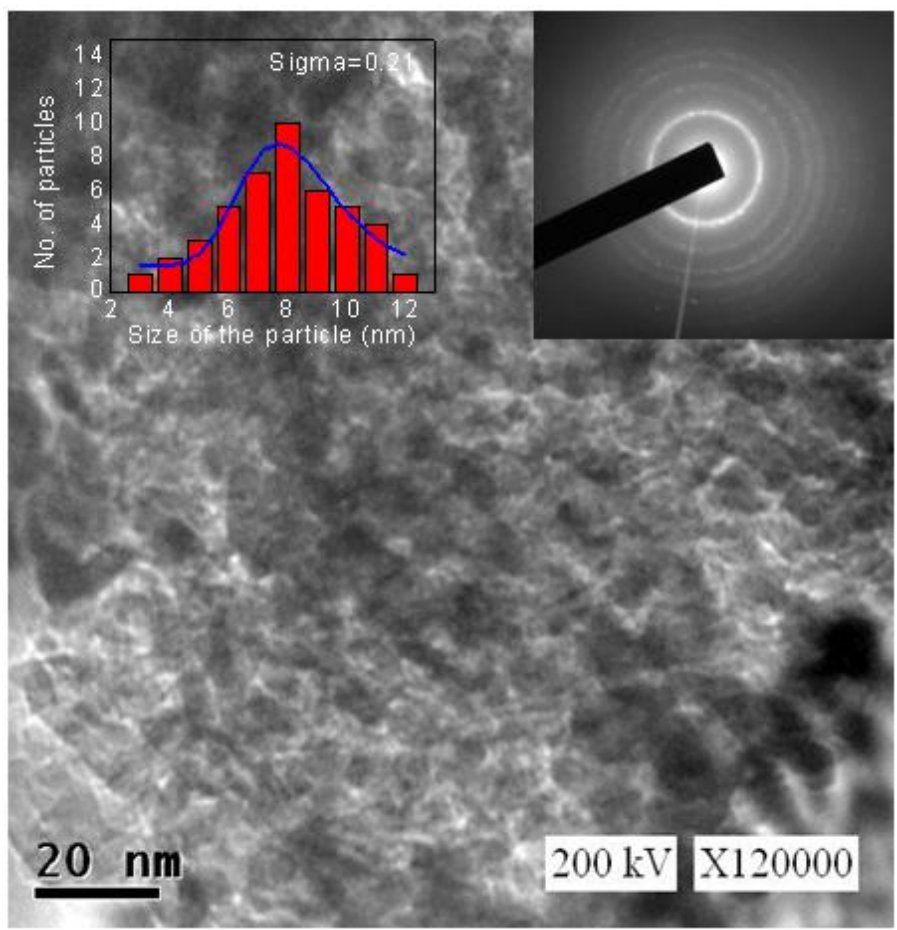

**Figure 3**

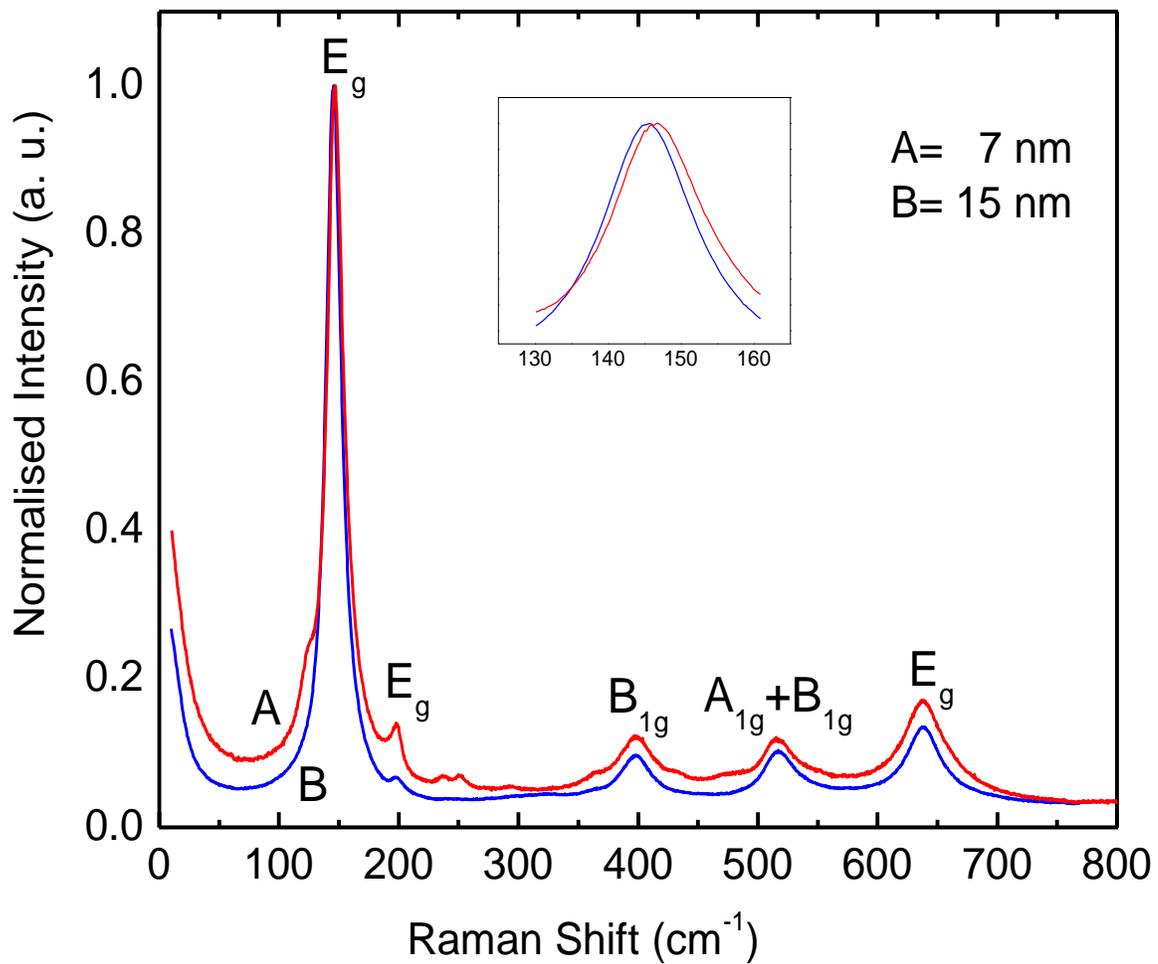

**Figure 4**

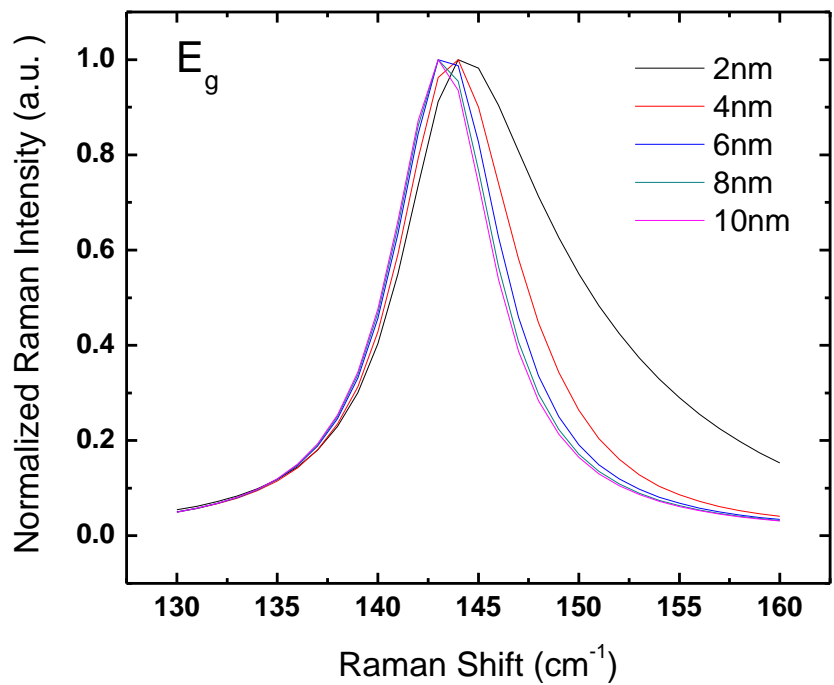

**Figure 5**

(a)

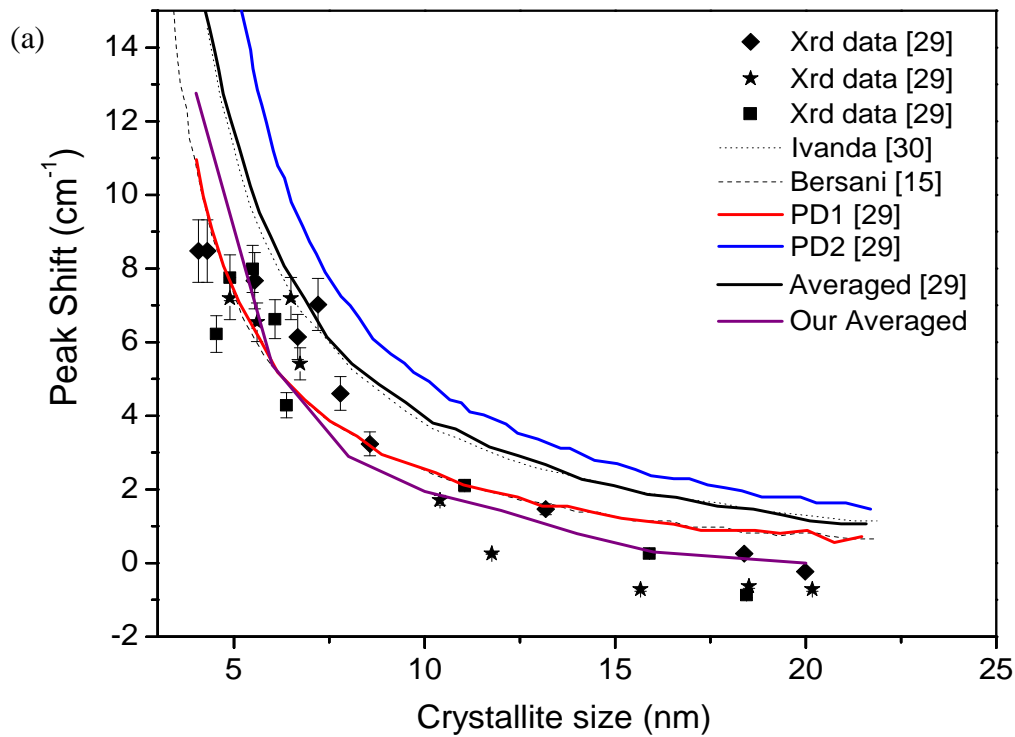

(b)

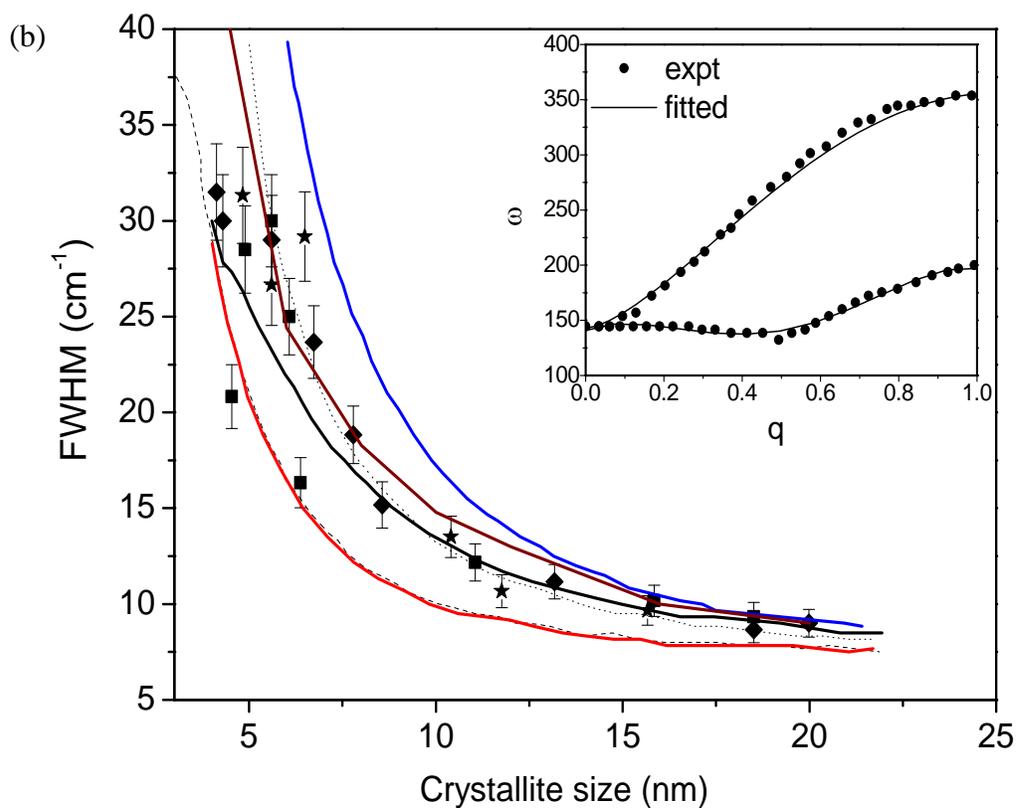

**Figure 6**

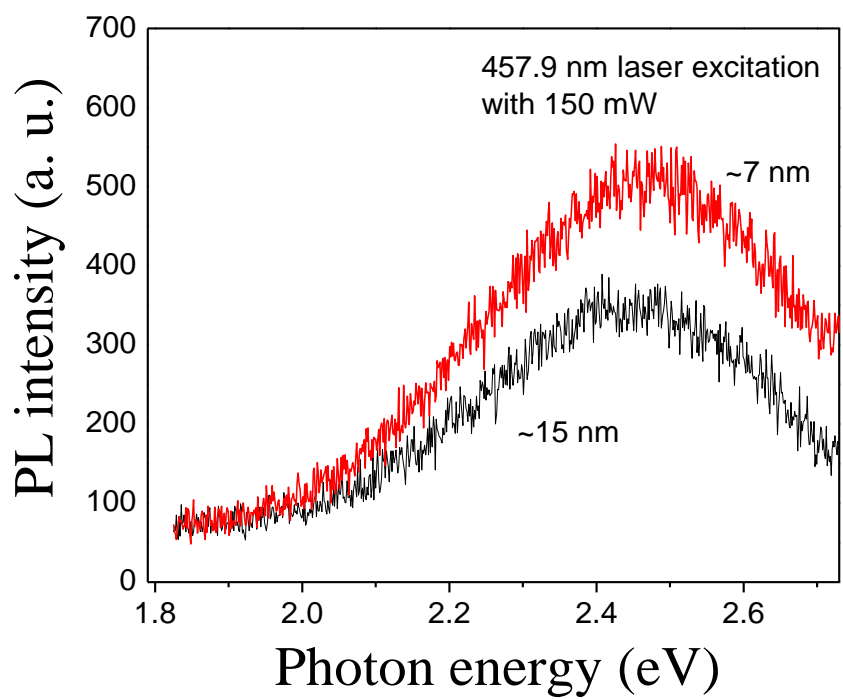